\documentclass[11pt]{article}
\usepackage{epsfig}

\newcommand{\anabla}{{\overrightarrow{\nabla}}\!\!\!\!\!\!{\overleftarrow{\nabla}}}

\newcommand{\del}{\partial}
\newcommand{\beq}{\begin{eqnarray}}
\newcommand{\eeq}{\end{eqnarray}}
\newcommand{\be}{\begin{eqnarray*}}
\newcommand{\ee}{\end{eqnarray*}}
\newcommand{\bk}{{\bf k}}
\newcommand{\bp}{{\bf p}}
\newcommand{\bq}{{\bf q}}

\newcommand{\ra}{\rightarrow}
\newcommand{\e}{\epsilon}

\begin{document}

\centerline{\Large\bf {Relativistic corrections to the Pionium Lifetime}}
\vskip 10mm
\centerline{Xinwei Kong$^1$ and Finn Ravndal\footnote{On leave of absence from Institute 
            of Physics, University of Oslo, N-0316 Oslo, Norway}} 
\medskip
\centerline{\it Department of Physics and Institute of Nuclear Theory,}
\centerline{\it University of Washington, Seattle, WA 98195, U.S.A}

\bigskip
\vskip 5mm
{\bf Abstract:} {\small Next to leading order contributions to the pionium lifetime are
considered within non-relativistic effective field theory. A more precise determination of the
coupling constants is then needed in order to be consistent with the relativistic 
$\pi - \pi$ scattering amplitude which can be obtained from chiral perturbation
theory. The relativistic correction is found to be 4.1\% and corresponds simply to a more
accurate value for the non-relativistic decay momentum.}


\bigskip

In the DIRAC experiment which is underway at CERN, one plans to measure the pionium 
lifetime with an accuracy of 10\% or better\cite{DIRAC}\cite{Schacher}. The dominant decay 
proceeds through the strong annihilation $\pi^+ + \pi^- \ra \pi^0 + \pi^0$. Since the
momenta of the final state particles is given by the mass difference $\Delta m = m_+ - m_0$
between the charged and neutral pions, the process is strongly non-relativistic. In lowest
order the decay rate follows directly from the corresponding non-relativistic 
scattering amplitude which can be written as
\beq
     T_{NR} = {8\pi\over 3E_+E_0}(a + b p^2/m_+^2)                                 \label{amp}
\eeq
in the center-of-mass frame where the energies of the pions are $E_0 = E_+ = m_+ + p^2/2m_+$ when
$p$ is the momentum of the charged pions. The S-wave scattering length $a$ and the slope parameter 
$b$ include both higher order chiral corrections and isospin-violating effects from the quark mass 
difference $m_u - m_d$ and short-range electromagnetic effects. At threshold the momentum $p=0$ 
and the full scattering amplitude with long-range Coulomb interactions removed is then given by
just this scattering length. It can be written in terms of the more conventional isospin-symmetric 
scattering lengths $a_0$ and $a_2$ in the isospin $I=0$ and isospin $I=2$ channels as $a = a_0 - 
a_2 + \Delta a$ where $\Delta a$ includes these symmetry-breaking effects. 

Since the charged pions in pionium are supposed to be bound in a
$1S$ Coulomb state $\Psi(r)$ with relative momentum $\gamma = \alpha m_+/2$, the annihilation 
takes place essentially at threshold. In lowest order the transition rate is then 
given by just the scattering length $a$,
\beq
     \Gamma = {16\pi\over 9m_+^4}|\Psi(0)|^2 m_0\sqrt{2\Delta m m_0}\, a^2
                                                                         \label{Gamma_0}
\eeq
where $|\Psi(0)|^2 = \gamma^3/\pi$ gives the probability to find the two particles at the same 
point. A measurement of the pionium lifetime will thus give an experimental determination of
the scattering length. For this to be meaningful, corrections to this lowest order formula
must be calculated so that the theoretical lifetime has an uncertainty substantially smaller
than the one in the experiment.

In a recent paper we have described this system in the new framework of non-relativistic effective
field theory\cite{KR}. This approach offers a compact and more systematic approach to obtaining
the many different higher order corrections to the pionium lifetime compared with previous
methods which to a large extent were based on covariant methods\cite{refs}. As shown by 
Holstein\cite{BH}  to lowest order in the interactions this approach is equivalent to just
using effective couplings in non-relativistic quantum mechanics.

The effective field theory for non-relativistic pions is based upon the free Schr\"odinger 
Lagrangian ${\cal L}_0 = \pi^\dagger(i\del_t + \nabla^2/2m_\pi)\pi$ which corresponds to the 
propagator
\beq
    G(E,\bk) = {1\over E - \bk^2/2m_\pi + i\e}                            \label{prop}
\eeq
for a particle with energy $E$ and momentum $\bk$. The interactions we consider are contained
in the  Lagrangian\footnote{We define here $C_0$ with opposite sign to what we used in \cite{KR}}
\beq
     {\cal L}_{int} = - {1\over 2}C_0 (\pi_+^* \pi_-^*\pi_0 \pi_0)
     + {1\over 4}C_2 (\pi_+^* \pi_-^*\pi_0 \anabla^2 \pi_0 + 
       \pi_+^* \anabla^2 \pi_-^*\pi_0 \pi_0) + \mbox{h.c.}                \label{int}
\eeq
where the gradient is defined as $\anabla = 1/2(\overrightarrow{\nabla} - \overleftarrow{\nabla})$.
It makes it possible to calculate the scattering amplitude for $\pi^+ + \pi^- \ra \pi^0 + \pi^0$ to
order $p^2$ which then must agree with the definition (\ref{amp}) to this order. The result
of this matching for the first coupling constant is then
\beq
      C_0 = {8\pi\over 3m_+^2}\left[a - (b-a){\Delta m\over m_+}\right]
\eeq
when  we  neglect smaller rescattering corrections\cite{KR}. Similarly, one finds for the derivative
coupling $C_2 = (8\pi/3m_+^4)(b-a)$. These values are more accurate than the ones used 
previously\cite{KR}.

With the value of the coupling $C_0$ now determined, one can calculate the second order correction 
$\Delta E$ to the ground state energy of pionium from the bound state diagram in Fig.1 as first 
pointed out by Labelle and Buckley\cite{Labelle}. It is found to
be imaginary with a resulting decay rate of $\Gamma = - 2\,\mbox{Im}\,\Delta E$. To lowest order
in the mass difference $\Delta m$ and ignoring the small binding energy, this gives the 
zero-order result (\ref{Gamma_0}).

\begin{figure}[htb]
 \begin{center}
  \epsfig{figure=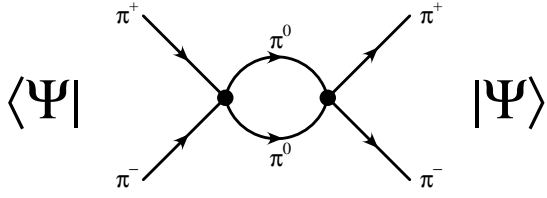,height=25mm}
 \end{center}
 \vspace{-8mm}            
 \caption{\small Next-to-leading order contribution to the ground state energy level shift which 
 gives  the leading order decay rate. The filled circle at the vertices denotes the $C_0$ coupling.}
 \label{fig1}
\end{figure}

Since the pion scattering lengths are set by the natural size $1/m_\pi$, the counting rules
needed to estimate the magnitudes of contributions appearing in different orders of
perturbations theory, are simple. The energy $E$ in the propagator will be of the order $Q^2$
when the characteristic momentum in the process is $Q$. As a result, the propagator scales as 
$1/Q^2$. For the same reason the four-dimensional volume integration $\int\!d^4k$ will scale as 
$Q^5$. The loop diagram in Fig. 1 thus scales as $Q$ since it involves two one-particle 
propagators. This is the leading order contribution to the decay rate.

To next order in the effective theory we must include the contribution from the derivative 
coupling $C_2$ in (\ref{int}). It will follow from the diagram in Fig.2 
and is seen to scale as $Q^3$ again using dimensional regularization\cite{KR}. The contributions 
to the decay rate from these two diagrams are thus found to be
\beq
    \Gamma = {m_0\over 4\pi} |\Psi(0)|^2 \sqrt{2\Delta m m_0}(C_0^2 + 2C_0C_2\Delta m m_0)
\eeq
With the above  value for the two coupling constants we then simply get the lowest order rate 
(\ref{Gamma_0}). The correction due to the slope parameter\cite{KR} thus disappears with the more 
precise values for the matched coupling constants used here. It results from a cancellation
between a next-to-leading order $p^2$ contribution and a $\Delta m$ contribution which in this
particular process are of the same order.
\begin{figure}[htb]
 \begin{center}
  \epsfig{figure=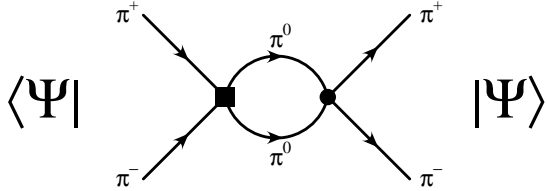,height=25mm}
 \end{center}
 \vspace{-8mm}            
 \caption{\small Next-to-next-leading order correction to the ground state energy level shift
which gives the next-to-leading order decay rate. The filled square  at the left vertex 
denotes the $C_2$ coupling.}
 \label{fig2}
\end{figure}

To this order in perturbation theory relativistic effects must also be included. These have 
previously also been considered in more covariant approaches\cite{refs}. Here they will arise
from the lowest order relativistic correction to the free Lagrangian ${\cal L}_0$ which now 
should be taken to be
\beq
      {\cal L} = \pi^\dagger \left(i\del_t + {\nabla^2\over 2m_\pi} 
                      + {\nabla^4\over 8 m_\pi^3}\right)\pi                   \label{L_rel}
\eeq
This new interaction will modify the propagators in the bubble of Fig.1 as shown in Fig.3.
\begin{figure}[htb]
 \begin{center}
  \epsfig{figure=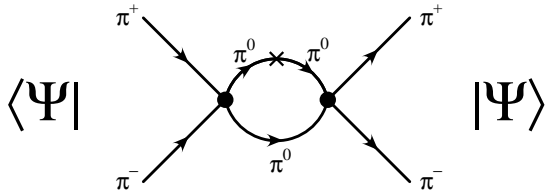,height=25mm}
 \end{center}
 \vspace{-8mm}            
 \caption{\small Relativistic correction to the decay rate. The cross denotes the 
relativistic coupling.}
 \label{fig3}
\end{figure}
Since the diagram now involves three propagators, it scales as $Q^5 Q^4 Q^{-6} = Q^3$.
This is also confirmed by an actual evaluation of the diagram which is given by
\beq
    \Delta E^{(rel)} =  \int\!{d^3p\over (2\pi)^3}\int\!{d^3k\over (2\pi)^3}
                       \int\!{d^3q\over (2\pi)^3} \Psi^*(\bp)\, C_0\, 
    {-\bk^4/8m_0^3\over (2\Delta m - \bk^2/m_0 + i\e)^2}\, C_0 \,\Psi(\bq)         \label{delta}
\eeq
Here $\Psi(\bp)$ is the Fourier transform of the bound state wavefunction. The integrations over 
momenta $\bp$ and $\bq$ now give just $|\Psi(0)|^2$. Using dimensional regularization for the 
divergent integral over the loop momentum $\bk$, we obtain the finite result
\beq
    \Delta E^{(rel)} = - {5i C_0^2\over 32\pi} \Delta m \sqrt{\Delta m m_0} |\Psi(0)|^2
\eeq
In terms of the decay rate, it corresponds to 
\beq
    {\Delta \Gamma^{(rel)} \over \Gamma} = {5\Delta m\over 4 m_+}         \label{Gamma_rel}
\eeq
and amounts to 4.1\%. By its very nature it can also be derived from using more covariant 
methods\cite{refs}. Finally, the same  relativistic interaction will also act on
the external legs of the diagram in Fig.1. But the resulting correction is then of the order
$\alpha^2$ and can thus be neglected here.

The relativistic correction (\ref{Gamma_rel}) follows also directly from the available
phase space for the $\pi^0\pi^0$ final state since there is no energy dependence in the 
annihilation amplitude to lowest order. Each $\pi^0$ has the energy $E_0 = m_0 
+ {k^2/ 2m_0} - {k^4/ 8m_0^3}$ when we include the next-to-leading order term in the momentum
expansion. The decay rate will then involve the integral
\beq
     \Gamma \propto \int\!{d^3k\over (2\pi)^3}\;\delta(2m_+ - 2E_0)
\eeq
when we ignore the small binding energy. The argument of the delta-function will now have two zeros
of which one represents an unphysical, high-momentum state. Keeping only the contribution from the
physical state, we recover exactly the additional term (\ref{Gamma_rel}).

In a recent paper by Gall, Gasser, Lyubovitskij and Rusetsky\cite{GGLR} the pionium lifetime
is also calculated from the non-relativistic Lagrangian used above. But instead of
using bound state perturbation theory based upon the standard Coulomb wavefunction as done here, 
they determine the properties of the bound state by calculating the complex pole on the second 
Riemann sheet of the corresponding T-matrix. In this way they derive higher order corrections 
which are not considered here. Our results are consistent with their general form of the decay rate
where the relativistic correction (\ref{Gamma_rel}) is seen to be the first term in the
expansion of their decay momentum $p^*$. The higher order corrections they have derived should 
follow in the present approach from including the Coulomb-interactions between the charged pions
and rescattering effects corresponding to extra bubbles in the diagram Fig.1.

Almost at the same time another calculation of the lifetime was completed by Eiras and 
Soto\cite{ES}. This is again based upon the same effective Lagrangian which is now further reduced
by integrating out degrees of freedom with momenta of the order $\sqrt{2\Delta mm_0}$. It then
becomes an effective theory for only charged pions with contact and Coulomb interactions. Their
result for the lifetime is also in agreement with what we have obtained here.

We want to thank P. Labelle for several helpful comments and J. Gasser, P. Labelle, 
V.E. Lyubovitskij, A. Rusetsky and J. Soto for many clarifying discussions during the 
HadAtom99 workshop. In addition, we are grateful to the Department of Physics and the INT 
at the University of Washington in Seattle for generous support and hospitality. Xinwei Kong 
was supported by the Research Council of Norway.

\end{document}